\documentclass[superscriptaddress,showkeys,showpacs,runinaddress,twoside,prb,twocolumn]{revtex4}

\usepackage{textcomp,graphicx}  

\newcommand{\scite}[1]{~\cite{#1}}

\usepackage{amsmath,amssymb}

\newcommand{\clead}{Chapter~}
\newcommand{\flead}{Figure~}

\newcommand{\cref}{\clead\ref}
 
\newcommand{\fref}{\flead\ref}

\newcommand{\dash}{~--- } \newcommand{\nbh}{\nobreakdash-}
\let\latin\textit \newcommand{\uprightlatin}{\let\latin\relax}
\newcommand{\ie}{\latin{i.e.}, } 
 \newcommand{\etal}{\latin{et al.}}

 \newcommand{\mAA}{\text\AA}

\newcommand{\bvect}[1]{{\mathbf #1}} 
\newcommand{\bhat}[1]{\hat{\bvect{#1}}} 
\let\vect\avect \let\uvect\ahat
\newcommand{\boldvectors}{\let\vect\bvect \let\uvect\bhat}

\newcommand{\flatpd}[2]{\partial{#1}/\partial{#2}}

\newcommand{\avg}[1]{\left\langle{#1}\right\rangle}

\newcommand{\um}{\textmu m}

\uprightlatin
\newcommand{\co}{(Color~online)~}

\newcommand{\T}{t_t}
\newcommand{\cvn}{S$_1$} \newcommand{\cvnt}{S$_2$} 
\newcommand{\vn}{Case~\cvn} \newcommand{\vnt}{Case~\cvnt}
\newcommand{\vt}{Case~T} \newcommand{\rv}[1]{Case~R$_{#1}$}
\newcommand{\vl}{1.96} \newcommand{\vm}{2.95} \newcommand{\vh}{3.93}
\newcommand{\nm}{\text{~nm}} \newcommand{\ps}{\text{~ps}}
\newcommand{\kps}{\text{~km/s}}
\newcommand{\tD}{$t_D$} \newcommand{\emd}{\avg{d_\text{min}}}
\newcommand{\rzf}{r_0(r,u_p)} \newcommand{\vf}{v(r,u_p)}

\renewcommand{\flead}{Fig.~}

\begin{document}

\newlength{\figwidth}
\setlength{\figwidth}{\linewidth}

\title{Sensitivity effects of void density and arrangement in a REBO high explosive}
\author{S.~Davis~Herring}
\email{herring@lanl.gov}
\affiliation{Theoretical~Division, Los~Alamos~National~Laboratory, Los~Alamos,~NM~~87545}
\affiliation{Department~of~Applied~Science, University~of~California,~Davis, Davis,~CA~~95616}

\author{Timothy~C.~Germann}
\email{tcg@lanl.gov}
\affiliation{Theoretical~Division, Los~Alamos~National~Laboratory, Los~Alamos,~NM~~87545}

\author{Niels~Gr\o nbech\nbh Jensen}
\email{ngjensen@ucdavis.edu}
\affiliation{Department~of~Applied~Science, University~of~California,~Davis, Davis,~CA~~95616}


\begin{abstract}
The shock response of two-dimensional model high explosive crystals with various arrangements of circular voids is explored.  We simulate a piston impact using molecular dynamics simulations with a Reactive Empirical Bond Order (REBO) model potential for a sub-micron, sub-ns exothermic reaction in a diatomic molecular solid.  In square lattices of voids (of equal size), reducing the size of the voids or increasing the porosity while holding the other parameter fixed causes the hotspots to consume the material more quickly and detonation to occur sooner and at lower piston velocities.  The early time behavior is seen to follow a very simple ignition and growth model.
The hotspots are seen to collectively develop a broad pressure wave (a sonic, diffuse deflagration front) that, upon merging with the lead shock, transforms it into a detonation.  The reaction yields produced by triangular lattices are not significantly different.  With random void arrangements, the mean time to detonation is 15.5\% larger than with the square lattice; the standard deviation of detonation delays is just 5.1\%.
\end{abstract}

\maketitle

\section{Introduction}
Heterogeneities such as inclusions, voids, cracks, and other defects enhance the shock sensitivity of high explosives by causing additional shock dissipation that creates small regions of high temperature called hotspots\scite{info:lanl-repo/inspec/1653210}.
Chemical reactions initiated in the hotspots emit pressure waves that merge with the lead shock and strengthen it, so that further hotspots are created with more vigor.  This positive feedback is the principal mechanism of the shock-to-detonation transition in inhomogeneous explosives\scite{beyond-standard}.
While the significance of heterogeneities is well known, which of their characteristics are most important are not.  In particular, since the details of the growth of reactions from the hotspots are not well understood, it is not known whether hotspots act separately or if the spatial arrangement of hotspots determines their efficacy.

Spherical voids are an often-studied, common defect in explosives\scite{zukas/walters,bowden/yoffe,explosives/propellants,beyond-standard}.  They can become hotspots upon collapsing under shock loading, and may also cause hotspots elsewhere by their partial reflection of the lead shock and by emitting further shocks upon their collapse and explosion.  Inert beads produce similar effects; while they do not collapse violently enough to become hotspots, their reflected shocks are not weakened by immediately following rarefactions and so may more easily create hotspots where they collide.

Experimental, theoretical, and numerical studies have sought to explain the sensitivity enhancement caused by voids and inert inclusions.
Bourne and Field\scite{info:lanl-repo/eixxml/1991060235317} reported results from shocked two-dimensional samples of gelatin or an emulsion explosive that had large cylindrical voids introduced.  They observed that voids could shield their downstream neighbors from the lead shock, but that they could also effect the collapse of their neighbors by emitting shock waves when they collapsed.
Dattelbaum \etal\scite{nm-hotspots2} shocked samples of nitromethane with randomly embedded glass beads or microballoons, observing that their presence decreased the run distance to detonation and the pressure dependence of that distance.  The balloons were found to have a greater effect than the beads, and small beads were in turn more effective than large beads.

Medvedev \etal\scite{emulsion-microballoons} conducted a theoretical analysis of emulsion explosives with microballoons that explained changes in detonation velocity with microballoon concentration via an ignition and growth model with a constant mass burn rate per hotspot.
Bourne and Milne\scite{cavity-collapse} experimentally and computationally considered a hexagonal lattice of cylindrical voids in an emulsion explosive or nitromethane and observed that the reactions at the hotspots accelerated the shock relative to a comparison with water.
In a previous report\scite{paper1} we used molecular dynamics (MD) to simulate single circular voids (and their periodic images) in a two-dimensional model solid explosive; the one rank of voids was able to induce a shock-to-detonation transition in the downstream material.
In this study, we extend those simulations to samples with structured and unstructured two-dimensional arrangements of voids.

\section{Method}
We simulate piston impacts on a number of samples with several equal-sized circular voids either randomly placed or in a regular square or triangular lattice.  We parameterize the possible lattices by their symmetry, the total porosity $p$ (proportion of molecules removed from the perfect crystal), and the radius $r$ of each void.  The spacing between voids in the square lattice is then
\begin{equation}
\delta\equiv r\sqrt{\pi/p}.
\end{equation}
The principal goal is to determine which of these parameters have a significant effect on the sensitivity of the explosive, as measured by the time \tD\ from piston impact to detonation transition.
We also look for nonadditive contributions from the arrangement of voids (rather than their simple number density) and explore the mechanism of the development of detonation.

\subsection{Model}
The Reactive Empirical Bond Order (REBO) ``AB'' potential (originally developed in~\cite{info:lanl-repo/inspec/4065531,info:lanl-repo/inspec/4411199,
extreme-dynamics}) describes an exothermic $\mathrm{2AB \rightarrow A_2 + B_2}$ reaction in a diatomic molecular solid and exhibits typical detonation properties but with a sub-micron, sub-ns reaction zone that is amenable to MD space and time scales.  Heim \etal\ modified it to give a more molecular (and less plasmalike) Chapman-Jouguet state\scite{info:lanl-repo/isi/000260573900087}.  We utilize the SPaSM (Scalable Parallel Short-range Molecular dynamics) code\scite{info:lanl-repo/eixxml/1994121436059} and the modified REBO potential (``ModelIV'') also utilized in our previous study~\cite{paper1}.  The masses of A and B atoms are 12 and 14~amu; a standard leapfrog-Verlet integrator is used with a fixed timestep of 0.509~fs in the NVE ensemble.

Our two-dimensional samples are rectangles of herringbone crystal with two AB molecules in each $6.19\times4.21$~\AA\ unit cell.  The shock propagates in the $+z$ direction; the samples are periodic in the transverse $x$ direction.  Each circular void is created by removing all dimers whose midpoints lie within a circle of a given radius.  The atoms are assigned random velocities corresponding to a temperature of 1~mK and an additional bulk velocity $v_z=-u_p$ directed into an infinite-mass piston formed by three frozen layers of unit cells at the $-z$ end.  The temperature is chosen to be small to avoid significant thermal expansion of the crystal but nonzero to avoid spurious effects from a mathematically perfect crystal; the RMS atomic displacement it causes is 2.1~pm.

Each simulation is run until the shock (whether reacting or not) reaches the free end of the sample; the traversal time of the shock (assuming that it does not accelerate) is $\T=Z/u_s(u_p)$, where $u_s(u_p)$ is the shock velocity Hugoniot.  It is broadly similar throughout the simulations, so the determination of whether or not a detonation occurs is meaningfully consistent.  In particular, in each of the principal studies the sample length $Z$ is held fixed so that $\T$ is constant for each $u_p$.
Analysis of the results, including dynamic identification of molecules, location of the lead shock, and determination of detonation transition times, follows~\cite{paper1}, except that when finding the detonation transition we keep the shock positions 15~nm apart for greater noise resistance.

\subsection{Systems}




We consider 94 square lattices: 27 with an integer number $n$ of voids (in each periodic image) and 67 with $n$ allowed to merely approximate an integer (so the last void's distance to the end of the sample differs slightly from the first's to the piston).  For brevity, we will term these two cases \cvn\ and \cvnt\ respectively.

In \vn, each combination of $p\in\{1.0,1.778,2.25\}\%$ and $r\in\{3,4,6\}\nm$ is considered, with $Z=1.28$~\um\ chosen so that $n\equiv Z/\delta$ is always an integer ($n\in[12,36]$), and each choice is simulated at each piston velocity $u_p\in\{\vl,\vm,\vh\}\kps$; 694776--2070048 atoms are simulated.  In \vnt, $Z=845\nm$ and $u_p=\vm\kps$ are fixed, and the 67 $p$-$r$ pairs from $\{1.0,1.1,\dotsc,2.0\}\%\times\{15,16,\dotsc,59\}~\mAA$ that yield an $n$ within 0.075 of an integer are simulated ($n\in[8,42]$, 260736--1341296 atoms).  The fixed velocity and loosened constraint on $n$ allow this study to explore the $p$-$r$ space more effectively.

In an ancillary study called \vt, rectangular and triangular lattices of $n=10$ voids are each simulated 27 times with a fixed $u_p=\vl\kps$ and every combination of $p\in\{1.0,1.5,2.0\}\%$, $r\in\{3,4,5\}\nm$, and $Z\in\{416,521,627\}\nm$.  Here the void spacings are $\delta_z=Z/n$ and $\delta_x=\pi r^2/p\delta_z=n\pi r^2/pZ$; 214060--1206248 atoms are simulated.

In the random case, a fixed sample size of $201\nm\times1015\nm$ is used with three different $(p,r,u_p)$ triples taken from the \vn\ lattices (but with more voids because of the increased sample area): \rv1 with $(1\%,6\nm,\vl\kps)$ and $n=18$, \rv2 with $(2.25\%,6\nm,\vl\kps)$ and $n=40$, and \rv3 with $(2.25\%,4\nm,\vm\kps)$ and $n=90$.  For each set of parameters, 10 simulations are run with different random arrangements of voids chosen by the simple rejection method such that all void centers are at least $2r$ away from either surface of the sample and at least $4r$ away from each other.  This largest case involves $\sim$3.1~million atoms.

\section{Results}
\begin{figure}
\centering\includegraphics[width=\figwidth]{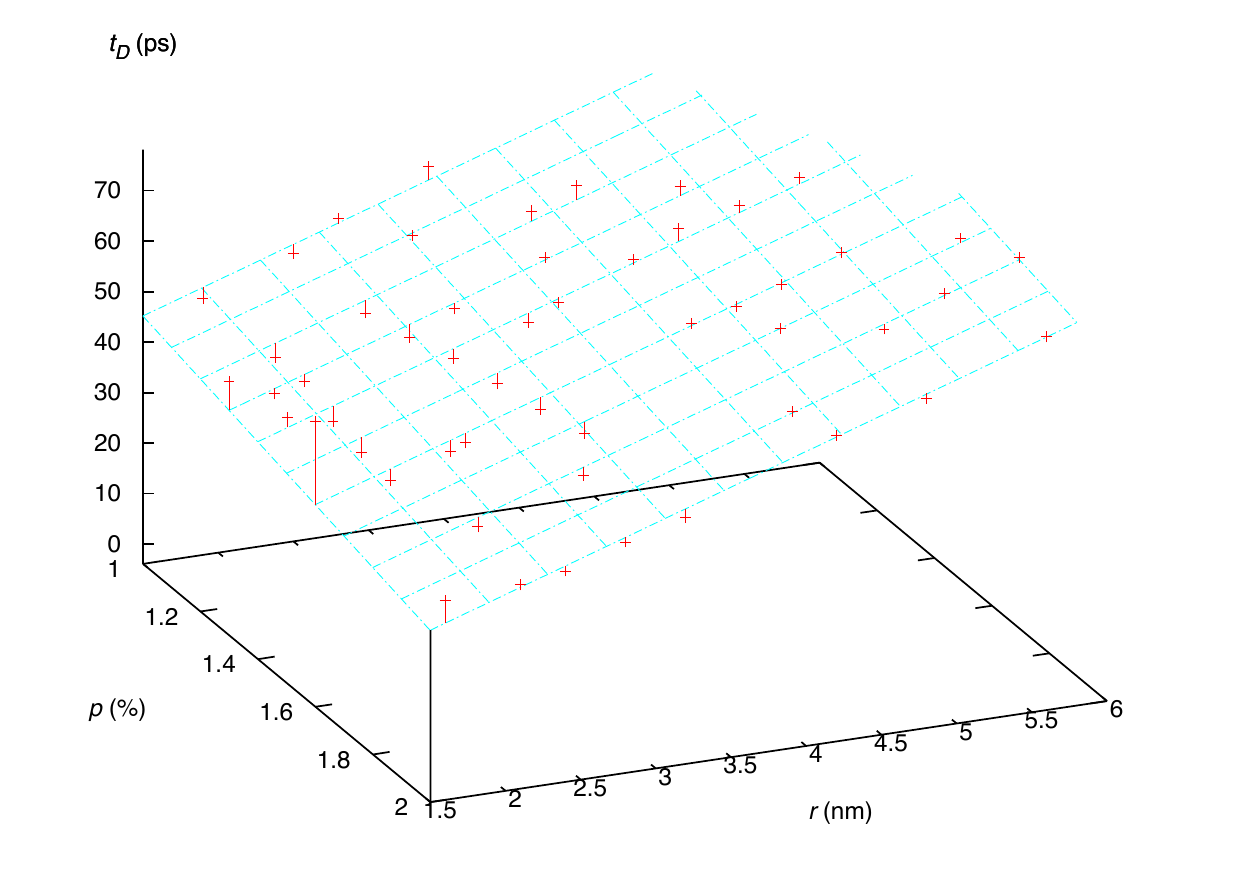}
\caption{Detonation transition times \tD\ from \vnt.  The grid shows every $p$ and every fourth $r$ value and is a planar fit to the data, which are shown as crosses connected to it by lines to place them in space and indicate the discrepancies.  The upper limit of the plotting region is the length of the longest, non-detonating simulations (78.2~ps).}
\label{vn2}
\end{figure}

The transition times for the 53 (of 67) \vnt\ samples that detonated are shown in \fref{vn2}.  The plane is a fit to the data; its equation is $t_D(p,r)/\text{ps}=9.718r/\text{nm}-1506p+45.67$.  The cases that did not detonate before the shock reached the free surface ($\T\approx78.2\ps$) would occupy the rear corner of the plot (smallest $p$ and largest $r$).  The proportion of atoms that form product molecules was generally 85\% but dropped to 65\% in that corner.  The half of those 14 simulations closer to the ones which detonated ended with detonation evidently imminent, but they are not counted as having detonated since a transition time could not be identified.

\begin{figure}
\centering\includegraphics[width=\figwidth]{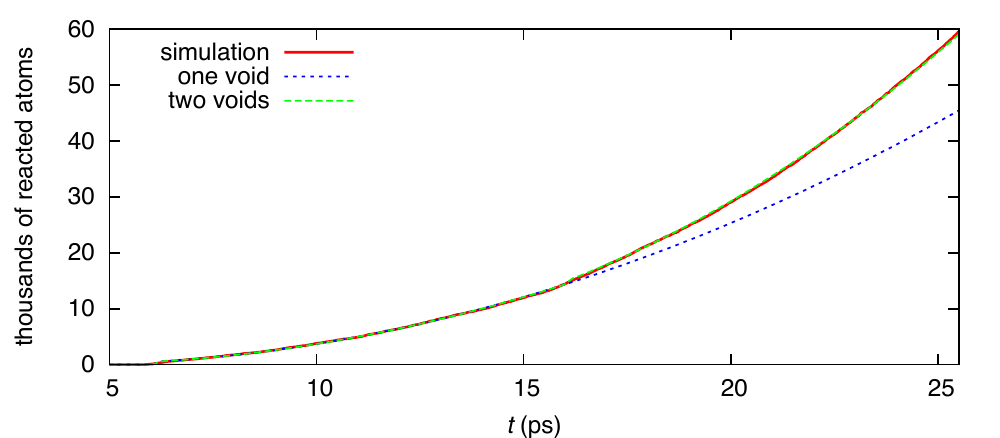}
\caption{Growth of reaction from the first two voids in a sample with $p=1\%$ and $r=59$~\AA.  The values from the ignition and growth model applied to the first void and to the first two voids are also shown; the latter curve is almost everywhere indistinguishable from the simulation values.}
\label{iandg}
\end{figure}

At early times, we observe that the extent of reaction closely follows a very simple ignition and growth model.  Suppose that a reacting hotspot is a disk that is created with a finite radius $r_0$ upon collapse of the void and grows at a constant speed $v$ into the surrounding unreacted material until it overlaps its periodic images.  If the disk contains a constant areal number density $n_a$ of reacted atoms, the number of reacted atoms as a function of time since initiation has the form
\begin{equation}
N(t)=n_aR(t)^2=n_a\pi(r_0+vt)^2.
\end{equation}
In \fref{iandg} are plotted the counts of reacted atoms from the beginning of the least reactive \vnt\ simulation, and the results of fitting one and two copies of $N(t)$ to them, corresponding to the first and then the second periodic line of voids being ignited.  The two copies use the same growth parameters and are merely each shifted in time to match the data.  Other \vnt\ simulations have similar behavior, but the growth parameters depend in an unknown fashion on $r$ and $u_p$, so we have no general model for $N(r,u_p,t)$.

\begin{figure}
\centering\includegraphics[width=\figwidth]{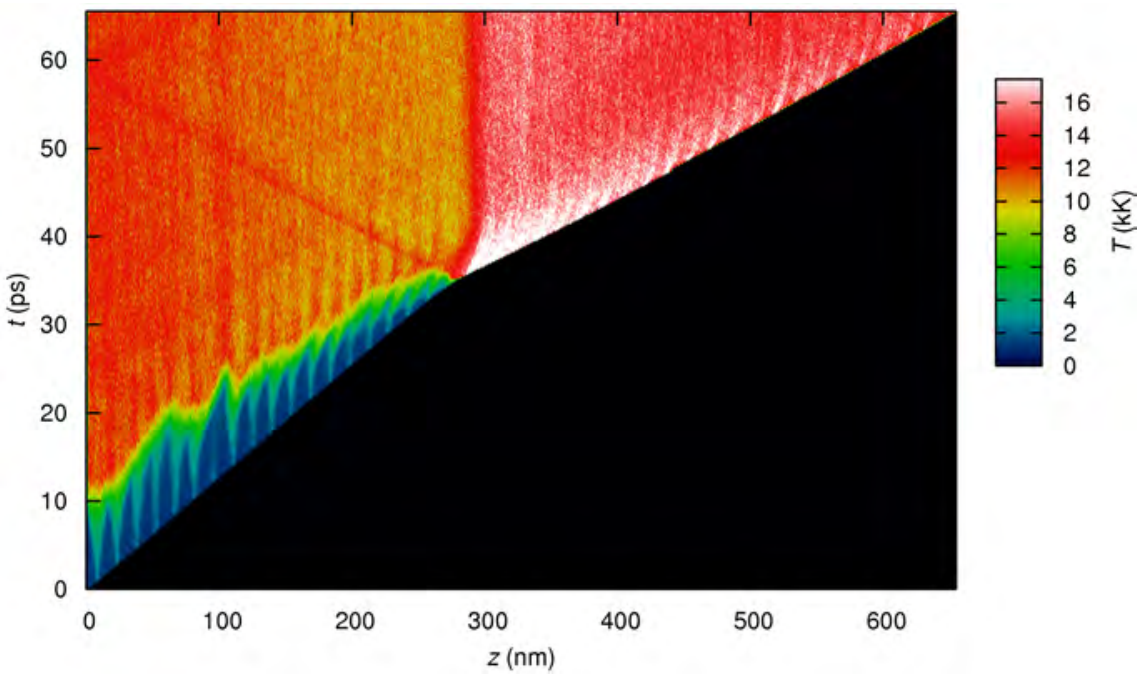}
\caption{\co Temperature over space and time in a sample with a square lattice of 42 voids: $p=2\%$, $r=16$~\AA, $u_p=\vm\kps$.  Only the region of space occupied by the sample at its final compression is shown.  Each spire at the lower left is a hotspot; the very hot region at large $z$, bounded below by a much faster shock, is the detonation.  Note the left-moving shock generated at the transition and the pressure wave (apparent as a temporary strong advection) visible in the shocked material between $z=150$~nm and $z=400$~nm.}
\label{spires}
\end{figure}

\fref{spires} shows the temperature distribution history from the other extreme \vnt\ simulation, with the smallest (eligible) $r$ at the largest $p$.  Each point in the figure is calculated with respect to the center of mass motion of, and averaged over, a column of computational cells of width $\Delta z\approx0.53$~nm.  We note here the development of a broad pressure wave in the shocked material, visible both as a region of strong advection intersecting the detonation transition and as a temperature increase between the last few hotspots created before the transition.  It appears that the pressure waves emitted by the first several hotspots merge and the combined wave strengthens itself by encouraging the deflagration at each hotspot it encounters.  When this wave overtakes the lead shock, its particle velocity is approximately equal to $u_p=\vm\kps$, so the relative velocity in the collisions at the shock doubles and detonation begins immediately.

\begin{figure}
\centering\includegraphics[width=\figwidth]{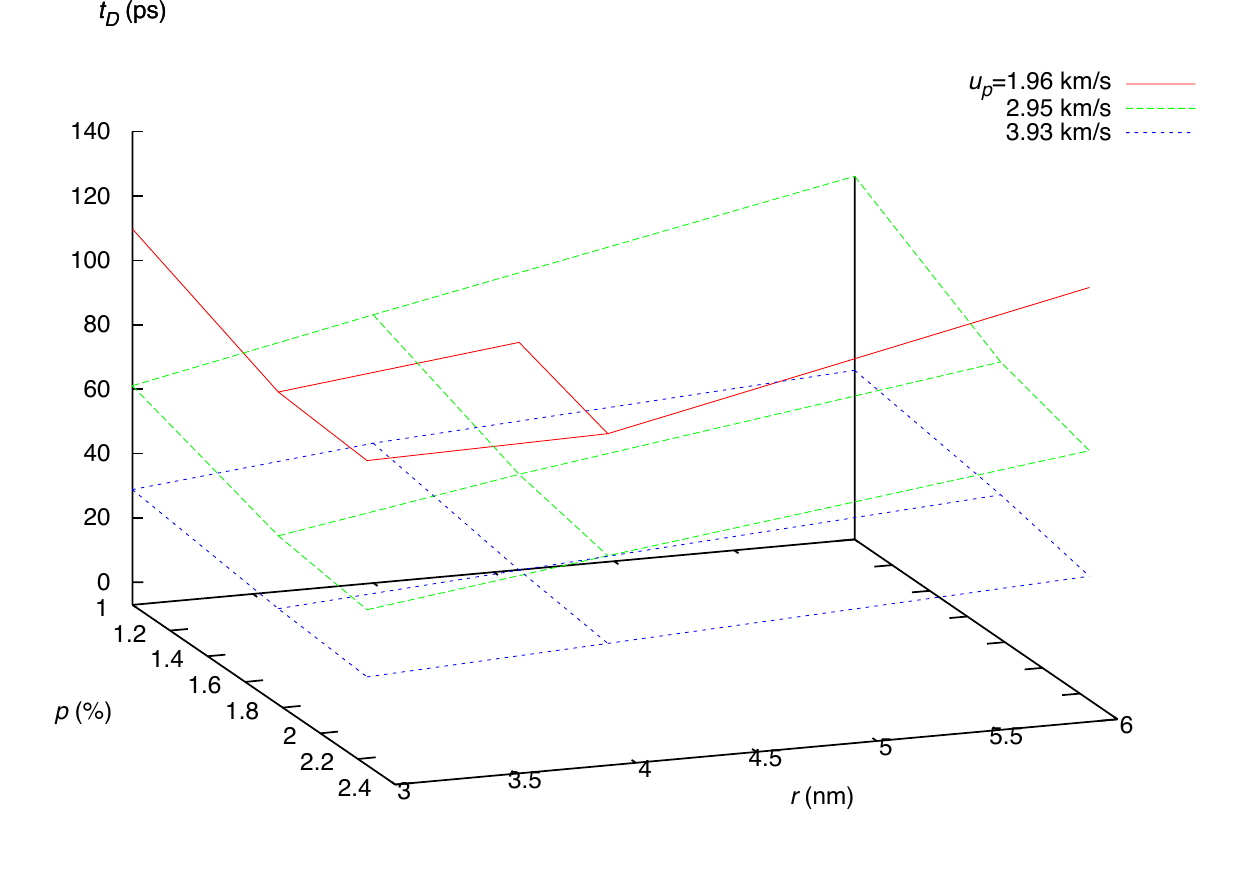}
\caption{Detonation transition times \tD\ from \vn.  The missing points for the smallest $u_p$ indicate failures to detonate.}
\label{vn}
\end{figure}

All but 3 of the 27 \vn\ simulations produced a detonation; those that did not used the lowest $u_p$ and (again) occupied the $-p$/$+r$ corner of that slice of the parameter space.  The transition times for the rest are shown in \fref{vn}; they follow the pattern of \vnt\ with the unsurprising addition that $\flatpd{t_D}{u_p}<0$.  The middle surface has the same $u_p$ as \vnt\ and shows some non-planarity beyond the range of \fref{vn2}.  With appropriate $p$ and $r$, we see detonations even at $u_p<2\kps$, which is much smaller than any value observed to trigger detonation with merely one periodic rank of voids\scite{paper1}; the feedback is much strengthened by the subsequent hotspots.

In \vt, the rectangular and triangular lattices did not differ significantly: they produced similar reaction yields and no detonations (presumably because they were relatively short; the \vn\ simulations corresponding to the most reactive case treated here detonated after 87--90~ps).  We would expect the triangular case to have a greater reactivity than the rectangular because the overlap between hotspots in adjacent columns is delayed by their offset.  The difference is never more than 5\% of the area, however, and lasts only until the overlap is complete (about 30\% of the reaction phase), so it may be difficult to measure.

\begin{figure}
\centering\includegraphics[width=\figwidth]{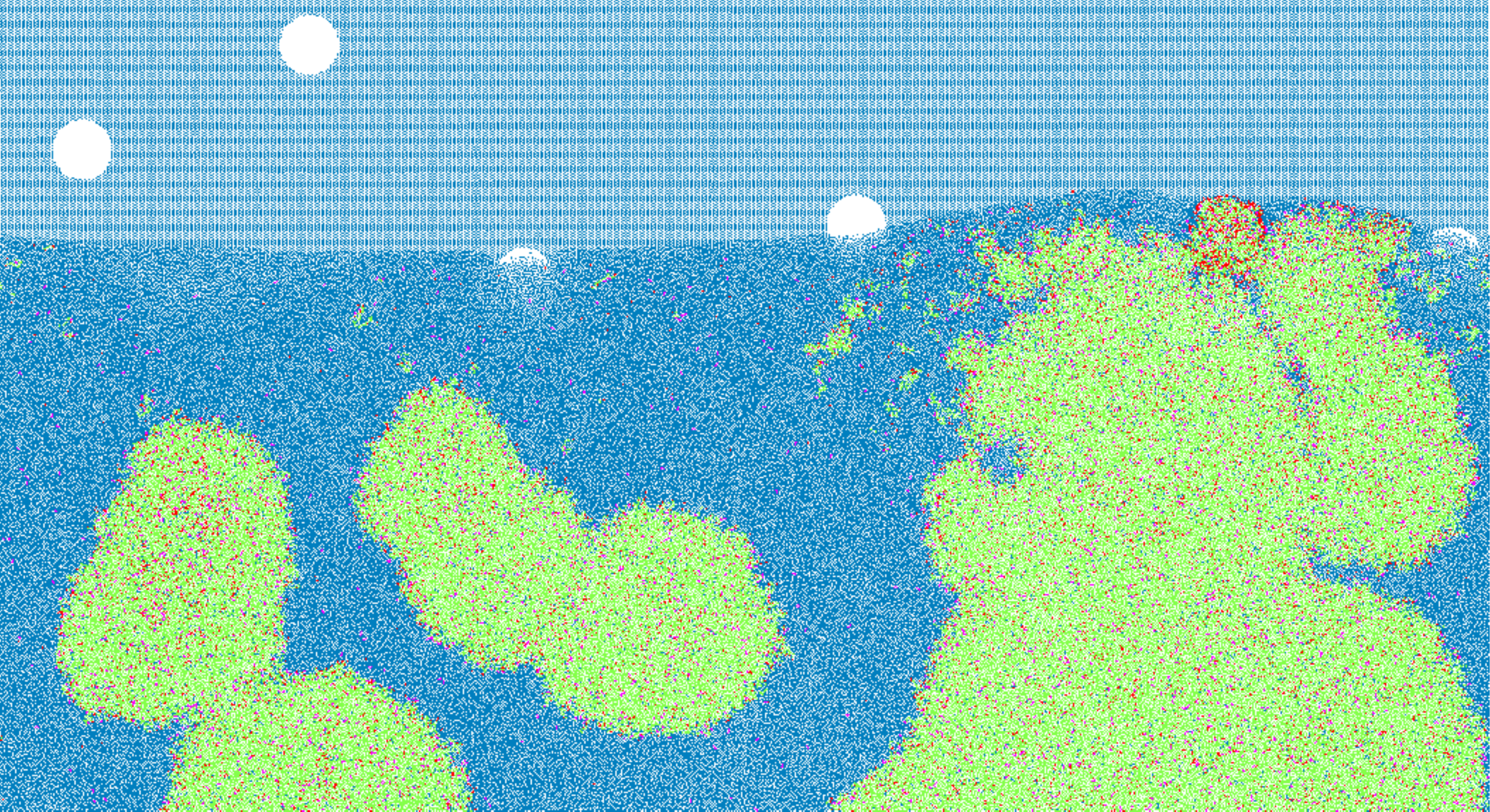}
\caption{\co Snapshot from a \rv3\ simulation just as the upward-propagating shock is becoming a detonation.  The whole width of the sample but only one ninth of its (original) length is shown.  Blue atoms are unreacted, green are reacted, and red and purple are intermediate states.  The red disk is a fresh hotspot; note the isolated reactions close to the shock everywhere.}
\label{rvrace}
\end{figure}

For \rv3, a consistent transition time of $t_D=59.3\ps\pm5.1\%$ was observed.  The mean is 15.5\% larger than the \tD\ from the corresponding \vn\ simulation; the small standard deviation suggests that the details of the void arrangement are not significant.
Furthermore, examination of one such simulation shows that a tight arrangement of 5 voids approximately 30\% of the way down the sample produces no extra reactions.  Later, a triangle of 6 voids triggers the transition, but spontaneous reactions elsewhere along the shock are also doing so simultaneously, as shown in \fref{rvrace}.
None of the other random arrangements detonated.  \rv2 produced yields of $68.0\%\pm3.7\%$ (where the second percentage is relative) whereas its \vn\ counterpart detonated.  \rv1 produced $52.1\%\pm6.3\%$ as compared to 70.1\% for its counterpart.  These differences are between normalized quantities, yet are partially due to the fact that the \vn\ samples were 26.1\% longer.

\section{Discussion}
It is to be expected that the detonation transition time \tD\ decreases with an increase in either the porosity $p$ or piston velocity $u_p$, but that it increases with radius ($\flatpd{t_D}r>0$) deserves further consideration.  First it should be noted that $(\flatpd{t_D}r)_\delta<0$: enlarging each of a set of voids without moving them (\ie keeping the separation $\delta$ fixed) does enhance the reactivity.  However, when holding $p$ fixed the reduction in number density overwhelms the effect of increasing the void size.

The simple ignition and growth model used earlier provides an explanation.  Before any detonation begins, any point in the material is reacted if and only if it is closer to the location of a void collapse than the $R(t)$ associated with that void\scite{info:lanl-repo/inspec/9166224}.  Since all voids near a given point will collapse at nearly the same time, what matters is the expected minimum distance $\emd$ to a void (after shock compression).  We expect that $\emd\propto\delta\propto r$, so the material will react sooner, on average, with small voids\dash until, of course, the voids become so small that they no longer reliably produce any reactions at all.  (For the smallest $u_p$ considered here, that failure radius is approximately 2~nm\scite{paper1}.)

A caveat is that if $r_0$ or $v$ is a strong function of $r$, the larger average distances from larger voids might be ovewhelmed by more vigorous growth of the hotspots created by larger voids.  While we do not have a model for $\rzf$ and $\vf$, they appear to be weak (perhaps sublinear) functions of $r$, in which case the conclusion of more reactivity from smaller voids holds.  That $v$ is a function of $r$ at all is interesting; we suppose that the $r$-dependent strength of the reshock emitted when the void collapses and explodes in place may imprint on its surrounds a memory of the void's size.

The model also explains how disorder in the arrangement of voids increases \tD.  Whenever, in the random placement of voids, two or more are placed much closer than $\delta$ to one another, their hotspots will overlap very quickly and the total burn front area will then be reduced; equivalently, $\emd$ is larger for a random arrangement than for a lattice (especially a hexagonal one) at the same $p$ and $r$.

The broad pressure wave created by the lead hotspots seems to be the principal mechanism for the detonation transition in this system.  Its development, the identical growth of the first two hotspots, the similarity of the results from rectangular and triangular lattices, the consistency among the results from random void arrangements, and the apparent irrelevance of void clusters all suggest that the development of a detonation is a collective effect that depends on $p$, $r$, and the regularity of the void arrangement but not on the details of that arrangement.  This collectivity affords a major simplification in predicting the behavior of collections of voids: a model might need only $r$, $p$, and $\sigma_\delta$.

Finally, we note that the function $\vf$, since it will likely dominate $\rzf$ and appears to depend on both its arguments but not on time, may prove useful as a measure of the strength or activity of a hotspot that might be incorporated into an analytical reaction rate model.  For voids of non-uniform size, it might be sufficient to consider the variation of $v$ in calculating a point's expected burn time.

\begin{acknowledgments}
This report was prepared by Los Alamos National Security under contract no.~DE-AC52-06NA25396 with the U.S. Department of Energy.  Funding was provided by the Advanced Simulation and Computing (ASC) program, LANL~MDI contract~75782-001-09, and the Fannie and John Hertz Foundation.
\end{acknowledgments}


\end{document}